\documentclass[a4paper,12pt]{article}

\usepackage{jcappub} 
\usepackage{textcomp}
\usepackage[T1]{fontenc} 
\usepackage{hyperref}
\usepackage{graphicx}
\usepackage{lscape}
\usepackage{array}
\usepackage[table,dvipsnames]{xcolor}
\usepackage{colortbl}

\title{Does Local Structure Within Shock-sheath and Magnetic Cloud Affect Cosmic Ray Decrease?}

\author[a,1]{Anil Raghav,\note{Corresponding author.}}
\author[a]{Zubair Shaikh,}
\author[b,1]{Ankush Bhaskar,}
\author[a]{Gauri Datar,}
\author[b]{Geeta Vichare}

\affiliation[a]{University Department of Physics, University of Mumbai,\\ Vidyanagari, Santacruz (E),
Mumbai-400098, India}
\affiliation[b]{Indian Institute of Geomagnetism,\\ New Panvel, Navi Mumbai-
410218, India.}

\emailAdd{raghavanil1984@gmail.com}
\emailAdd{ankushbhaskar@gmail.com}
\emailAdd{}

\abstract{The sudden short duration decrease in cosmic ray flux is known as Forbush decrease  which is mainly caused by interplanetary disturbances. A generally accepted view is that the first step of Forbush decrease is due to shock sheath and second step is due to magnetic cloud (MC) of interplanetary coronal mass ejection (ICME). This simplistic picture does not consider several physical aspects, such as, whether the complete shock-sheath or MC (or only part of these) are contributing to the decrease, what effect does the internal structure within the shock-sheath region / MC have on the decrease, etc. We present a summary of the analysis of a total of 18 large ($\geq8 \%$) Forbush decrease events and the associated ICMEs, a majority of which show multiple steps in the Forbush decrease profile. We propose a re-classification of Forbush decrease events depending upon the number of steps observed in their respective profile, and the physical origin of these steps. Our analysis clearly indicates that not only broad regions (shock-sheath and MC), but also localized structures within the shock-sheath and MC, have a very significant role in influencing the Forbush decrease profile.  The detailed analysis in the present work  is expected to contribute toward understanding the relationship between FD and ICME parameters in better way.

\textit{\textbf{key words}}: Shock-sheath, magnetic cloud (MC), ICME, cosmic ray, Forbush decrease, local magnetic structures. 
}

\begin{document}
\maketitle
\flushbottom

\section{Introduction}

Sudden short duration decrease in cosmic ray count observed on the surface of Earth by neutron monitor (NM), followed by recovery time lasting from $\sim$ few hours to days and sometime even weeks, are known as Forbush decrease \cite{Forbush1937}.  These decreases are typically caused by interplanetary counterparts of coronal mass ejections (ICMEs) which  generally consist of `shock-sheath'  and `Magnetic Cloud (MC)'~\cite{Cane2000,Anil2014}. The modulation of cosmic ray flux occurs due to their interaction with the convected interplanetary magnetic field. The study of these cosmic ray modulations might lead to unravel the structure of interplanetary disturbances (e.g. ICME shock, MC).

The ICME can be thought of as a closed, low density magnetic tongue-like structure that expands as it propagates through the interplanetary space. Its higher speed relative to the background solar wind speed produces a shock-sheath region ahead of it. This magnetic field has an ordered, as well as a turbulent component. The magnetic field of the CME is assumed to be frozen into the plasma. As CME takes-off, initially it is devoid of cosmic ray particles, however cosmic rays diffuse into ICME during its propagation. In spite of this diffusion, the interior of the ICME is still deficient in cosmic rays as compared to its surroundings. Therefore, when the ICME passes through the Earth, a decrease in intensity of cosmic rays is observed. This is how a Forbush Decrease takes place \cite{Richardson2010, Richardson2011, Vourlidas2013}. On the basis of the above model/understanding, Forbush decrease events are classified in two categories: (i) one-step Forbush decrease and (ii) two-step Forbush decrease. The observation of one-step or two-step Forbush decrease depends not only on the structure of the interplanetary disturbance, but also on the location of the observer. Due to the presence of two regions (shock-sheath and MC), the general nature of Forbush decrease is of two-step type. However, if the observer crosses only shock-sheath or only MC then it manifests as a one-step decrease. One-step decrease is also observed when the ICME is less energetic (and hence without significant shock-sheath). Hence, broadly, there are three categories of Forbush decrease, which are caused by  (i) only shock (one-step), (ii) only MC (one-step) or (iii) a combination of shock and MC (two-step). Further, on an average roughly equal contribution of the shock and MC to the overall decrease in the Forbush decrease is observed \cite{Richardson1996,Wibberenz1998}. 

The role of ICME in Forbush decrease events has been studied intensively by many researchers around the globe (e.g. \cite{ Richardson1996, Cane2000, Richardson2011,Belov2001, Belov2008, Belov2014, Dumbovic2011, Dumbovic2012, Papaioannou2010}). In spite of all these number of attempts to relate Forbush decrease properties with measured parameters of ICMEs at 1 AU, there are significant gaps in our understanding of their underlying physical mechanisms. This could be due to the Forbush decrease being a manifestation of large scale interplanetary disturbances, whereas the spacecraft measurements are local and might not reflect the global property of the disturbance \cite{Richardson2011}. These studies indicate that interplanetary magnetic field and solar wind speed affect Forbush decrease significantly through cross-field diffusion and convection-diffusion respectively (e.g \cite{Richardson1996, Wibberenz1998, Bhaskar2016}). The significant contribution of the turbulent sheath region of the ICME in determining the Forbush decrease magnitude has been pointed out previously \cite{Badruddin2002, Badruddin2002a, Anil2014,Subramanian2009, Candia2004, Giacalone1999}. However, Jordan \textit{et al}. (2011), propose rejection of the two-step Forbush decrease model and emphasize the necessity to include the contribution of ignored small-scale interplanetary magnetic structure~\cite{Jordan2011}. In the light of this proposal, it becomes necessary to re-examine the `traditional' understanding of Forbush decrease in detail.

As presented in the above discussion, it is well-understood that the ICME is the real contributor in the non-recurrent Forbush decrease. The literature suggests that the first step of two-step Forbush decrease is due to shock-sheath and second step is due to magnetic cloud of ICME. However, there are a few questions which are important to address: (i) Does the complete ICME shock or MC (or only some part of them) contribute in cosmic ray decrease?  (ii) Is there any internal structure associated with the ICME shock / MC which is, in particular, responsible for the decrease? Therefore, in this work, we carry out a detailed study of each of 18 strong Forbush decrease ($\geq8\%$) events and associated ICMEs, occurred in the last two decades. Here, we present our findings after closely inspecting the structural aspect of ICME and its corresponding Forbush decrease.

\section{Data and methodology}
We have used  the  five  minute  resolution  neutron  flux  data  from  51 neutron monitor (NM) observatories. Each NM observatory is located at different altitude, latitudes and longitudes and having different vertical geomagnetic cut-off (\textit{i.e.} rigidities), the detailed information is available online, on \url{www.nmdb.eu} and \url{http://nest2.nmdb.eu/} data sites. The 18 Forbush decrease events have been selected for investigation from year 1998 to 2011 in the Forbush decrease catalogue given on the same database. The list of events is mentioned in Table~[\ref{Table : 1}]. 
Though Forbush decrease is a global phenomenon, individual observatories will have different baselines due to their respective instrumental characteristics and local parameters. To minimize these variations, we normalized the cosmic ray intensity of each observatory. The  normalized percentage variation (\%) for each NM observatories is defined as 
\begin{equation}
{{N_{norm}(t)}={\frac{N(t)-N_{mean}}{N_{mean}}}\times{100}}                           
\end{equation}

where  $N_ {mean}$  is average of previous day/days (quiet time) counts of a particular observatory and  $N(t)$ is the neutron count at time $t$ of the same observatory. We have used 5-minute time resolution interplanetary data taken  from the OMNI database \url{http://cdaweb.gsfc.nasa.gov/cgi-bin/eval1.cgi} and ACE database (4 min average IMF data and 64 sec average solar wind data)  \url{http://www.srl.caltech.edu/ACE/ASC/level2/}. 

The amplitude and profile of the observed Forbush decrease generally vary with the energy of the cosmic rays. Moreover, the neutron monitors are distributed globally with varying geomagnetic cut-off rigidity which provide an opportunity to study Forbush decrease in different energy bands. Therefore, to study general energy-wise variation we have categorized neutron monitor data for each event into three categories: (i) low rigidity (0 -- 2 GV), (ii) Medium rigidity ( 2 -- 4.5 GV ) and  (iii) high rigidity (> 4.5 GV). Further, the average global response is studied using the mean neutron flux variation.  Along with neutron flux, the associated interplanetary parameters of ICMEs are studied in detail. 

To identify the ICME components, researchers use various criteria. Generally, the following criteria are used to identify MC: (i) enhanced IMF, (ii) slowly decreasing in solar wind velocity, (iii) low proton density, (iv) low plasma temperature. Whereas, to identify shock-sheath, the following criteria are used: (i) enhanced IMF and solar wind speed, (ii) high proton density, (iii) high plasma temperature. We have utilized the boundaries of shock-sheath and MC using available at  \url{space.ustc.edu.cn/dreams/wind_icmes/} and the paper by Richardson and Cane (2010)~\cite{Richardson2010}. In all the figures presented here, the onset of the shock is shown by the first vertical black dashed line and the MC region is represented by the dashed rectangular magenta box. The Forbush decrease onset is determined by using visual inspection where neutron flux starts to decrease sharply. For every Forbush decrease event, each step in Forbush decrease profile is indicated by black arrow-heads.

\section{Observations \& Discussion}
 
To understand broad features of the Forbush decrease profile, the studied events are classified based on the number of steps observed in cosmic ray flux corresponding to shock-sheath and MC. The details of these events are presented in Table~[\ref{Table : 1}]. The following subsections discuss a representative event of each category in detail. The remaining events are shown in the attached supplementary document. 

\subsection{One step classical Forbush decrease profile}
\begin{figure}[!ht]
\includegraphics[width=1.0\textwidth]{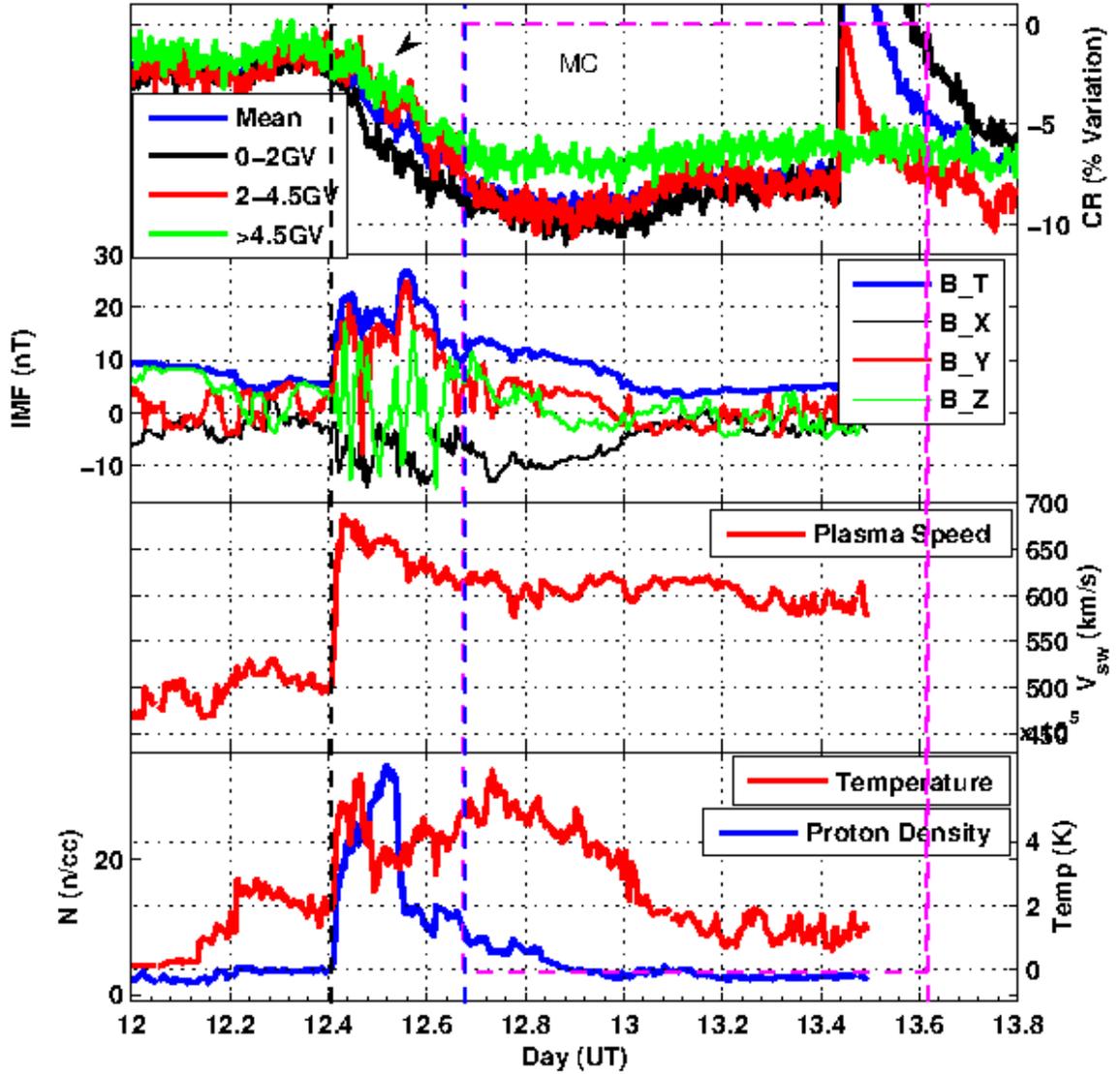}
 \caption[Forbush decrease event occurred on  July 13, 2000 ]{\textit{Forbush decrease event occurred on July 13, 2000. It has four panels, top most panel shows temporal variation of normalized neutron flux with their respective band of rigidities. The arrow shows step decrease in cosmic ray. The $2^{nd}$ and $3^{rd}$ panels show interplanetary magnetic field ($B_{Total}$ and $B_X$,$B_Y$,$B_Z$-component) and solar wind speed data respectively. The bottom panel shows proton density and plasma temperature variation.  }}
 \label{fig:3}
 \end{figure}
The events occurred on July 13, 2000 and February 18, 2011 represent one step classical Forbush decrease profile i.e. the cosmic ray flux shows single step decrease during ICME transit. Here, Figure [\ref{fig:3}] describes the representative event which occurred on July 13, 2000. The figure clearly shows that the onset of Forbush decrease observed in all neutron monitors is simultaneous with the arrival of interplanetary shock-sheath at the Earth\textquotesingle s Bow-shock nose. The arrival of shock can be identified by the sharp enhancement in total IMF B and solar wind speed. The shock-sheath region lasted for $\sim 6$ hours (bounded by two dashed vertical lines from left). The overall gradual decrease is observed in the cosmic ray flux for all energies (top panel). 
The cosmic ray flux decrease occurring during shock-sheath transit  suggests that the first step gradual decrease in Forbush decrease is caused due to the shock-sheath component of ICME. It is important to note that the IMF components mainly $B_z$ show fluctuations throughout the shock-sheath region. This might imply the presence of turbulent magnetic field inside the sheath which causes the gradual decrease in cosmic ray.  

As the Earth entered in MC, the neutron monitor flux showed gradual variation reaching minimum and then starts recovering within the MC. Note that IMF B, $V_{sw}$ and plasma temperature have steadily decreased and plasma density remained almost steady except some variations. Also, there is  no transient decrease observed in MC. 

Similarly, Forbush decrease event occurred on February 18, 2011 shows single step profile (Refer attached supplementary document). However, the shock-sheath associated with this event shows turbulent region just behind the shock-front which could have given rise to the observed decrease. During the later part of the sheath cosmic ray flux starts recovering. Interestingly, during MC the cosmic ray flux did not show further gradual/transient decrease, on the contrary it continued the gradual recovery.

\subsection{Two step classical Forbush decrease profile}
We have observed five classical two-steps Forbush decrease events. These events show first step transient decrease in cosmic ray flux during shock-sheath and second step in MC. These events include Forbush decreases occurred on September 17, 2000; January 22, 2004; January 21, 2005; May 15, 2005, September 11, 2005 and December 14, 2006.

   \begin{figure}[!ht]
\includegraphics[width=1.0\textwidth]{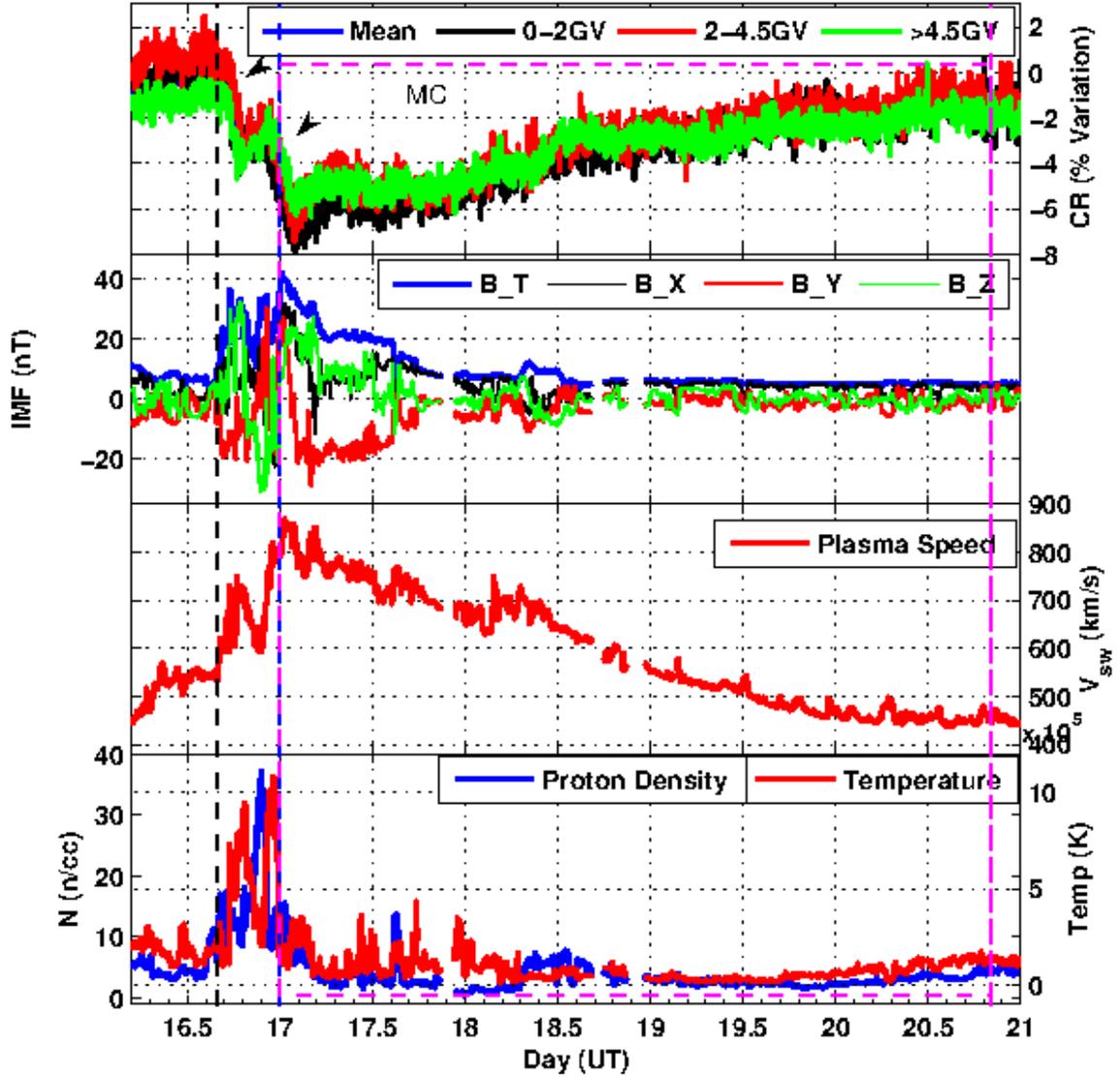}
 \caption[Forbush decrease event occurred on September 17, 2000]{\textit{Forbush decrease event occurred on September 17, 2000. It has four panels, top most panel shows temporal variation of normalized neutron flux with their respective band of rigidities. The arrow shows step decrease in cosmic ray. The $2^{nd}$ and $3^{rd}$ panels show interplanetary magnetic field ($B_{Total}$ and $B_X$,$B_Y$,$B_Z$-component) and solar wind speed data respectively. The bottom panel shows proton density and plasma temperature variation.}}
 \label{fig:5}
 \end{figure}
 
Figure \ref{fig:5} describes the representative two-step classical Forbush decrease event occurred on the September 17, 2000. The figure clearly shows that the onset of first step decrease observed in all neutron monitors is simultaneous with the arrival of interplanetary shock. The shock-sheath region lasted for $\sim 8$ hours. However, the first step decrease in cosmic ray is observed only for  $\sim 3.6$ hours after the onset and later it shows recovery.
This suggests that the first step decrease in Forbush decrease is caused by the shock-sheath component of ICME.  

The second step decrease started few hours before crossing the end part of the shock-sheath and minimum is observed during the leading edge crossing of MC. This could be interpreted as the transition region between shock-sheath and MC which significantly contributed in the second step of Forbush decrease. During the crossing of remaining MC, gradual recovery in cosmic ray flux is observed. 

Note that the total IMF shows enhancement corresponding to each step decrease. Also sharp fluctuations in y and z-components of IMF are observed. This can be interpreted as the presence of turbulence behind the shock-front and in the transition region. 
All other events under this category show similar features (Refer attached supplementary document).

\subsection{Multi-step complex Forbush decrease profile}
Along with classical one step and two step Forbush decrease profile, we have also observed multi-step complex Forbush decrease profile. Therefore, it is needed to introduce the extension of old Forbush decrease classification based on the number of steps decrease observed in Forbush decrease profile during shock-sheath and MC transit. 

\subsubsection{Multi-step Forbush decrease profile in shock-sheath (with one or no step in MC)}
 \begin{figure}[!ht]
\includegraphics[width=1.0\textwidth]{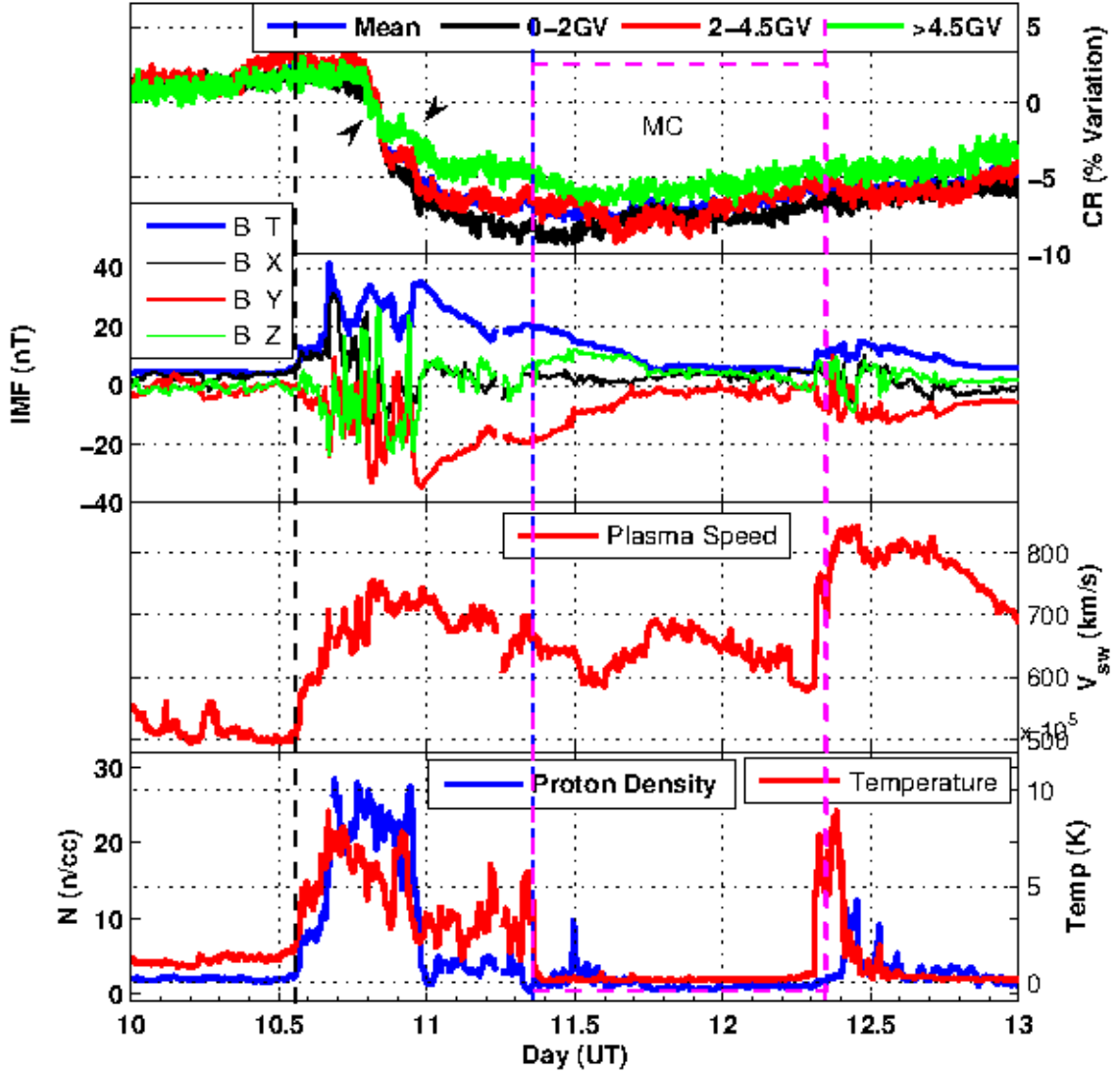}
 \caption[Forbush decrease event occurred on April 11, 2001]{\textit{Forbush decrease event occurred on April 11, 2001. It has four panels, top most panel shows temporal variation of normalized neutron flux with their respective band of rigidities. The arrow shows step decrease in cosmic ray. The $2^{nd}$ and $3^{rd}$ panels show interplanetary magnetic field ($B_{Total}$ and $B_X$,$B_Y$,$B_Z$-component) and solar wind speed data respectively. The bottom panel shows proton density and plasma temperature variation.}}
 \label{fig:6}
 \end{figure}
 
The complex Forbush decrease events showing more than one step in shock-sheath region and maximum one step during MC crossing  are studied here. 
We have observed total two Forbush decrease events under this category occurred on April 11, 2001 and September 25, 2001.  

Figure \ref{fig:6} describes a typical Forbush decrease event occurred on April 11, 2001. The shock-sheath region lasted for quite long $\sim 19.7 $ hours. Interestingly, the onset of Forbush decrease was delayed by $\sim 5.5 $ hours with respect to the shock arrival at the Earth.  Note that, observed IMF and all its components are highly fluctuating during the decrease as compared to the variations at shock-front. After the end of first step a small recovery is observed in cosmic ray flux. The second step decrease which occurred within shock-sheath followed this recovery, well before the Earth's entry into the MC. Corresponding to this second step the z-component of IMF shows sharp change and y-component of IMF shows maximum negative value. Rest of the shock-sheath shows gradual decrease in cosmic ray flux.  Also, corresponding IMF and its components show gradual variation which is exceptional within shock-sheath. Explicitly, y-component of IMF gradually recovers to its ambient value, whereas no significant variations are observed in x- and z-components of IMF. 

It is important to note that initial part of the shock sheath is less fluctuating, the intermediate part is highly fluctuating and the remaining part is ordered structure (seen as gradual variation in y-component of IMF) which later continued as MC. One needs to investigate the cause of this ordered structure seen prior to the MC. The similar multi-step features are seen in remaining events which are  shown in attached supplementary document.

\subsubsection{Multi-step Forbush decrease profile in MC (with one or no step in shock-sheath)}
   \begin{figure}[!ht]
\includegraphics[width=1.0\textwidth]{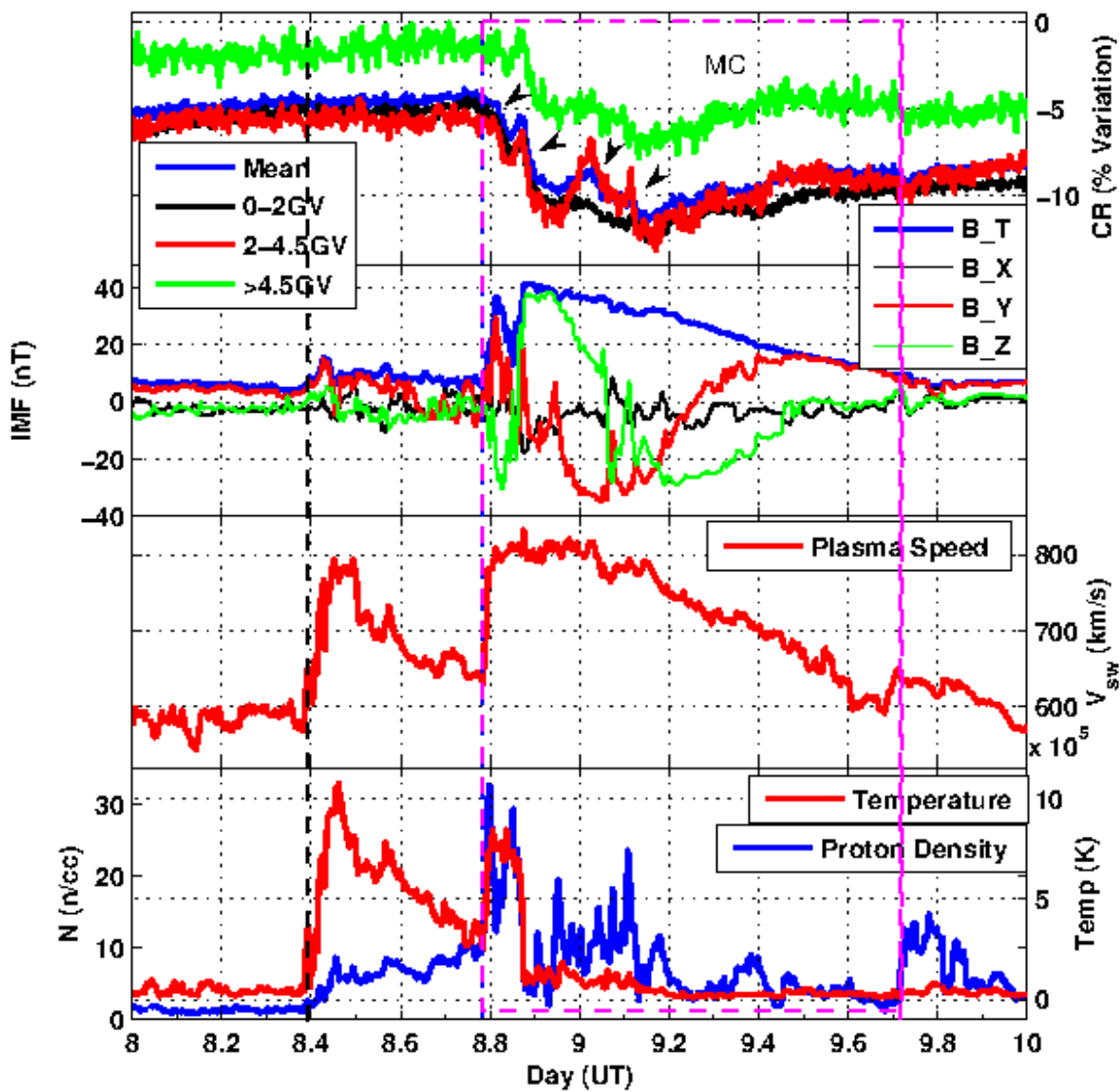}
 \caption[Forbush decrease event occurred on November 09, 2004]{\textit{Forbush decrease event occurred on November 09, 2004. It has four panels, top most panel shows temporal variation of normalized neutron flux with their respective band of rigidities. The arrow shows step decrease in cosmic ray. The $2^{nd}$ and $3^{rd}$ panels show interplanetary magnetic field ($B_{Total}$ and $B_X$,$B_Y$,$B_Z$-component) and solar wind speed data respectively. The bottom panel shows proton density and plasma temperature variation.}}
 \label{fig:13}
 \end{figure}

The Forbush decrease events showing maximum one step in shock-sheath region and multi-step decreases during  MC transit are considered here. We have observed total three Forbush decrease events under this category occurred on October 29, 2003; July 26, 2004 and November 09, 2004 respectively. 

The prototype event which occurred on November 09, 2004 is described in Figure [\ref{fig:13}]. For this event there is no decrease observed within the shock-sheath. Note that during passage of the shock-sheath, IMF and its components show less fluctuations. This could be responsible for the absence of cosmic ray decrease within the shock-sheath. 

As the Earth entered into MC, neutron monitors showed multiple sharp decreases in cosmic ray. However, there is no sharp variations in total IMF. The close inspection of the variations observed in IMF components reveals that there are sharp variations in  $B_y$ and $B_z$ components. This implies that the local spatial structures  within the MC have given rise to the multi-step decrease for this event. 

The similar multi-step features are seen in remaining events which are  shown in attached supplementary document.

\subsubsection{Multi-step Forbush decrease profile in shock-sheath and MC}
  \begin{figure}[!ht]
\includegraphics[width=1.0\textwidth]{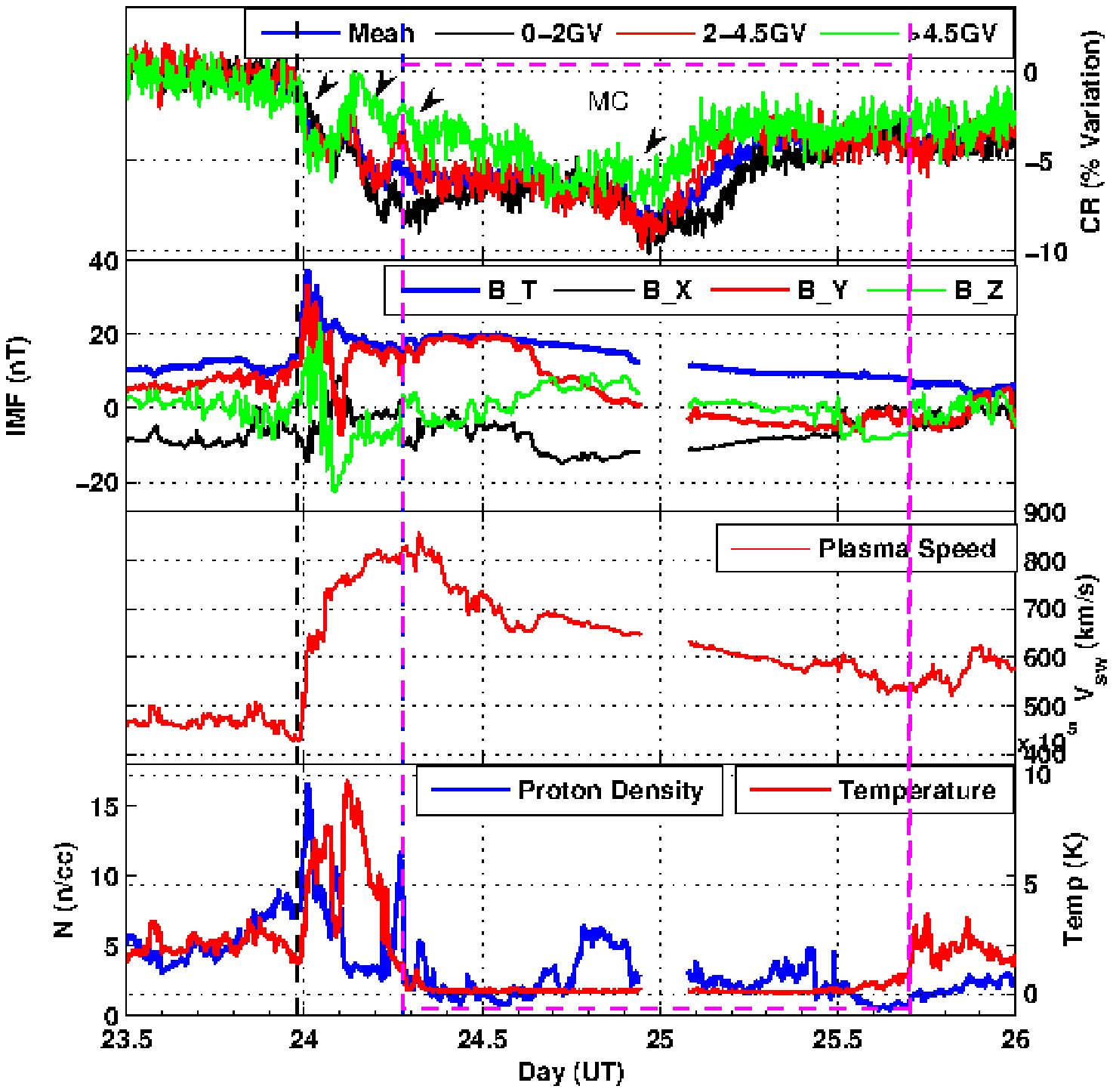}
 \caption[Forbush decrease event occurred on September 24, 1998]{\textit{Forbush decrease event occurred on September 24, 1998. It has four panels, top most panel shows temporal variation of normalized neutron flux with their respective band of rigidities. The arrow shows step decrease in cosmic ray. The $2^{nd}$ and $3^{rd}$ panels show interplanetary magnetic field ($B_{Total}$ and $B_X$,$B_Y$,$B_Z$-component) and solar wind speed data respectively. The bottom panel shows proton density and plasma temperature variation.}}
 \label{fig:2}
 \end{figure}
  
Forbush decrease events showing multiple step decreases in shock-sheath as well as in MC are investigated here. We have observed total five events occurred on September 24, 1998; July 15, 2000; November 06, 2001; November 24, 2001 and January 17, 2005
  
The typical event occurred on  September 24, 1998 is shown in Figure [\ref{fig:2}]. There are two sharp decrease in shock-sheath and two more in the MC. The figure clearly shows that the onset of Forbush decrease observed in all neutron monitors is simultaneous with the arrival of interplanetary shock at the Earth. The shock-sheath region lasted for $\sim 6$ hours. It is observed that the sharp recovery (especially in high energy cosmic rays) in neutron flux occurs  after $\sim 2.5$ hours of the onset. The larger recovery in high energy NMs compared to low energy NMs is  intriguing.
After the sharp recovery in neutron flux, once again it starts decreasing. During this second step decrease IMF B and its components remained steady and IMF was mainly directed in y-direction i.e dawn to dusk. Note that these both decrease occurred during shock-sheath. Just prior to the Earth entry into the MC there is a sharp recovery in cosmic ray flux. However, there is no sharp variation in IMF corresponding to this.

As the Earth enters in the MC neutron monitor flux shows minor sharp decrease and later remained low. Suddenly near day 25 there is a sharp decrease and then rapid recovery of cosmic ray counts is observed in all NMs. Unfortunately the interplanetary data is missing during this period, so that we are unable to see the interplanetary signatures associated with this. The similar multi-step features related to shock-sheath and MC are observed in other events of this category (Refer attached supplementary document ).

\section{Discussion \& conclusion}

The transient variations observed in cosmic ray flux at the Earth which are known as Forbush decrease are understood as the diffusion of cosmic ray through the ordered and turbulent large scale interplanetary magnetic field. The highly asymmetric (\textit{i.e.} sharp decrease and gradual recovery) and strong Forbush decreases are generally caused by ICMEs. These ICMEs have mainly two components: shock-sheath (assumed to be turbulent) and magnetic cloud (ordered structure) \cite{Richardson2011, Vourlidas2013}. The generally accepted view is that the classical two-step Forbush decrease has first step due to shock-sheath and second one due to MC \cite{Anil2014}. However, which part of shock-sheath and MC give rise to these steps is not well resolved yet. Therefore, understanding the role of local spatial structures within shock-sheath and MC during Forbush decrease motivated the present study.

To address this issue, we have investigated total 18 Forbush decrease events of large magnitude. These events are classified based on the observed number of steps in shock-sheath and MC. During shock-sheath transit we have observed following key features (summarize in Figure ~\ref{fig:6}) of the Forbush decrease profiles: (i) no step, one step or multi-step decrease in cosmic ray flux, (ii) simultaneous/non-simultaneous decrease with respect to the shock-front arrival at the Earth, (iii) gradual decrease throughout the transit of shock-sheath or short-duration sharp decrease corresponding to a small part of the shock-sheath. 

The absence of cosmic ray decrease during shock-sheath transit suggests the presence of weak/less turbulent shock-sheath. The one-step simultaneous Forbush decrease indicates the strong turbulent shock-sheath, however, non-simultaneous Forbush decrease indicates weak shock-front with localized turbulent sheath. The gradual decrease observed during complete shock-sheath transit implies the presence of turbulence throughout the shock-sheath. Whereas, short duration sharp decrease  corresponding to small part of the shock-sheath indicates the presence of turbulence in local region of shock-sheath. The multi-step decrease in cosmic ray could be due to the existence of multiple turbulent-regions in the shock-sheath.  

\begin{figure}[!ht]
\includegraphics[width=1.0\textwidth]{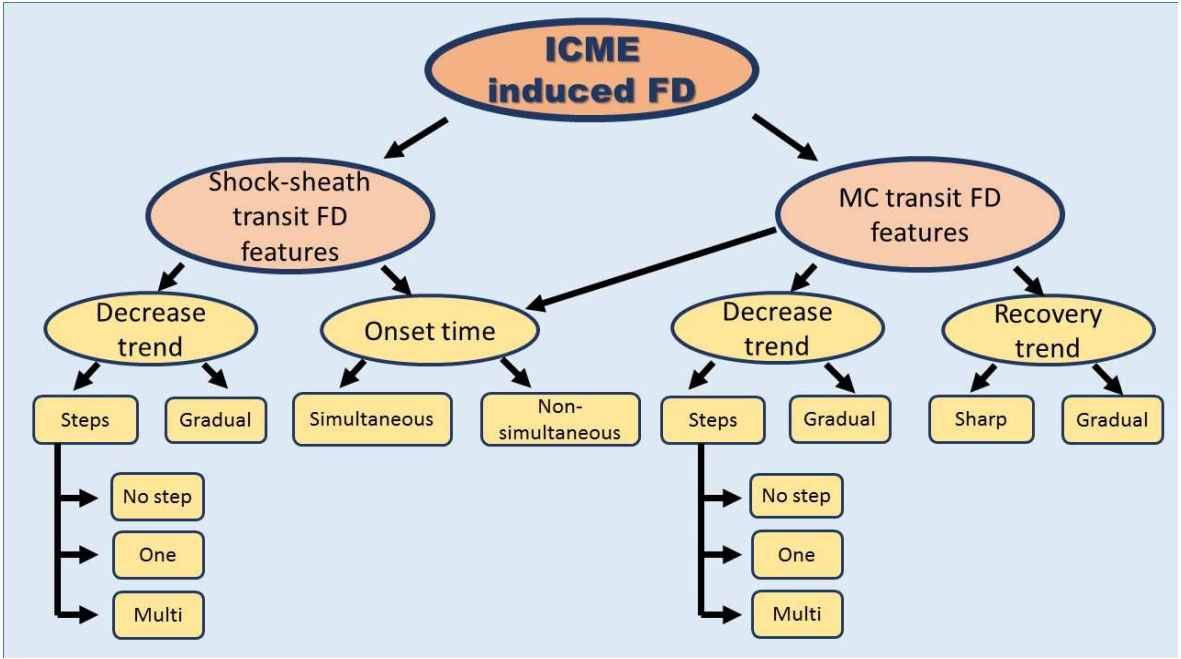}
 \caption[General features of Forbush decrease profile caused by ICME]{\textit{Summary of general features of Forbush decrease profile caused by ICME transit. The shock sheath and MC of ICME is considered to be the real driver of Forbush decrease profile in traditional model. However, the key features observed during the shock-sheath and MC transit (as summarize in above figures) are not explainable by the traditional model. These features demonstrate the significant role of localized structures within shock-sheath and MC in Forbush decrease profile.}}
 \label{fig:6}
 \end{figure}

This study also put some light on the role of MC in Forbush decrease profile. The one step decrease occurred during the crossing of shock-sheath end and front part of MC, implies the residence of turbulence just ahead of MC. However, most of the events show gradual decrease  or nearly constant cosmic ray flux during MC transit. This could be attributed to the inhibited diffusion of cosmic rays within MC due to ordered magnetic field. Also, nearly all events show recovery of cosmic ray within MC. This could be ascribed to the declining total IMF during the transit of MC. As the IMF is decreasing, the diffusion of cosmic rays into the MC increases which results in the recovery of cosmic ray flux. The multi-step decreases observed during MC transit could be due to the localized distortions of ordered magnetic structure within MC. There is a need of detail investigation of these local distortions and understand their origin.

The current understanding of Forbush decrease separately considers the contribution of shock-sheath and MC in Forbush decrease profile. Jordan et al.(2011) challenged this understanding and suggested the influence of small-scale interplanetary magnetic field structure in Forbush decrease profile. They concluded that traditional model of Forbush decreases having one or two steps should be discarded \cite{Jordan2011}. 
But, our observations of one-step and two-step Forbush decrease profile suggest that the framework of the traditional model (which is quantified by Raghav \textit{et al.} (2014)~\cite{Anil2014}) can be retained. As one can see the, observations of multi-step Forbush decrease profile (summarized features shown in Figure~\ref{fig:6}) cannot be explained by the traditional model only. This observations suggest the need of improved classification of Forbush decrease events. Hence, we  introduce a possibly better classification scheme for Forbush decrease events, as outlined in Table~[\ref{Table : 1}]. Also, the presence of multi-step decrease demonstrates the significant role of localized structures within shock-sheath and MC in Forbush decrease profile. 

In  summary, not only  broad regions (shock-sheath and MC) contribute to Forbush decrease but also localized structures within shock-sheath and MC have very much significant importance in Forbush decrease profile.
Therefore, one needs to be cautious about these local structures while studying relationship between Forbush decrease and ICME parameters. Also, accounting the influence of these local structures is essential for accurate modeling and understanding of cosmic ray transport in the heliosphere.

\section{Acknowledgement}
We acknowledge the NMDB database (www.nmdb.eu) founded under the European Union's FP7 programme (contract no. 213007). We are also thankful to all neutron monitor observatories listed on website. We are thankful to CDAWeb and ACE science center for making interplanetary data available. We are thankful to Department of Physics (Autonomous), University of Mumbai, for providing us facilities for fulfillment of this work. Authors would also like to thank S. Kasthurirangan for valuable discussion.

 \begin{table}
    \centering
    \caption[List of Forbush decrease events with classification]{\textit{ List of analysed Forbush decrease events classified on the basis of number of steps observed in cosmic ray flux.}}
    \vspace{0cm}
    \renewcommand{\arraystretch}{1.25}
    \setlength{\tabcolsep}{3pt}
    {\setlength{\extrarowheight}{5pt}}%
    \begin{tabular}{|m{5cm}|m{3.5cm}|m{3.5cm}|}
      \hline
       \hline
      
      \textbf{Event Date}&\multicolumn{2}{c|}{\textbf{No.of Steps in  cosmic ray Flux }} \\
         \cline{2-3}
       & \textbf{SS} & \textbf{MC}\\[2ex]
      \hline 
   
       \multicolumn{3}{c}{\textbf{(1) One step classical Forbush decrease profile}}  \\ [2ex]
       \hline 
       13 July 2000 &  1 & 0   \\ \hline
       18 Feb 2011 & 1 & 0 \\  \hline
       
       \multicolumn{3}{c}{\textbf{(2) Two step classical Forbush decrease profile}}  \\ [2ex]
       \hline
       17 Sept 2000 &  1 & 1  \\ \hline
       22 Jan 2004 & 1 & 1   \\ \hline
       21 Jan 2005 & 1 & 1   \\ \hline
       15 May 2005 & 1 & 1   \\ \hline
       11 Sept 2005 & 1 & 1   \\ \hline
       14 Dec 2006 & 1 & 1   \\ \hline
       \multicolumn{3}{c}{\textbf{(3) Multi-step complex Forbush decrease profile)}}  \\ [2ex]
       \multicolumn{3}{c}{(i) Multistep Forbush decrease profile in shock-sheath (with 1 or 0 step in MC)}  \\ [2ex]
       \hline
       11 Apr 2001  &  2  & 0    \\ \hline
       25 Sept 2001 & 2 & 0  \\ \hline
            
       \multicolumn{3}{c}{(ii)Multistep Forbush decrease profile in MC (with 1 or 0 step in shock-sheath)}  \\ [2ex]
       \hline
       29 Oct 2003 & 1 & 3  \\ \hline
       26 July 2004 & 1 & 2  \\ \hline
       09 Nov 2004 & 0 & 2    \\ \hline 
       
       \multicolumn{3}{c}{(iii)Multistep Forbush decrease profile in shock-sheath as well as in MC}  \\ [2ex]
       \hline
       24 Sept 1998 & 2  & 2  \\ \hline
       15 July 2000 & 2  & 2   \\ \hline
       06 Nov 2001  & 2 & 2  \\ \hline
       24 Nov 2001 & 2 & 2   \\ \hline
       17 Jan 2005 & 3 & 2   \\ \hline
       
   \end{tabular}
    \label{Table : 1}
  \end{table}
 \end{document}